\documentclass[pra,10pt,twocolumn,superscriptaddress,showpacs]{revtex4-1}
\usepackage{amsmath}
\usepackage{latexsym}
\usepackage{amssymb}
\usepackage{graphicx}
\usepackage[colorlinks=true, citecolor=blue, urlcolor=blue]{hyperref}
\usepackage{float}
\usepackage{amsfonts}
\usepackage{textcomp}

\begin{document}
\title{Entropy as a bound for expectation values and variances of a general quantum mechanical observable}
\author{Shubhayan Sarkar\\{\it School of Physical Sciences, National Institute of Science Education and Research,\\ HBNI, Jatni
- 752050, India}}
\date{August 6, 2017}

\begin{abstract}
Quantum information-theoretic approach has been identified as a way to understand the foundations of quantum mechanics as early as 1950 due to Shannon. However there hasn't been enough advancement or rigorous development of the subject. In the following paper we try to find relationship between a general quantum mechanical observable and {\it von Neumann entropy}. We find that the expectation values and the uncertainties of the observables have bounds which depend on the entropy. The results also show that {\it von Neumann entropy} is not just the uncertainty of the state but also encompasses the information about expectation values and uncertainties of any observable which depends on the observers choice for a particular measurement. Also a reverese uncertainty relation is derived for n quantum mechanical observables.\\
\\
{\bf Keywords:} Information-theoretic; von Neumann entropy; expectation values; variance; observables; uncertainty relation 
\\
\\
{\bf PACS numbers:} 03.67.-a; 03.65.-w; 89.70.Cf
\end{abstract}

\maketitle

\section{Introduction}
\label{intro}
Information theory resides at the heart of quantum mechanics. Information theory as a way to understand quantum mechanics has been a long sought out problem as early as 1956 due to Shannon \cite{8}. The idea was further refined by Hartle \cite{1}, a supporter of the {\it psi-epistemic} \cite{3} veiw of quantum mechanics. According to him \\
\\
{\it...A quantum mechanical state being a summary of the observers’ information
about an individual physical system changes both by dynamical laws, and
whenever the observer acquires new information about the system through
the process of measurement. The existence of two laws for the evolution
of the state vector...becomes problematical only if it is believed that the
state vector is an objective property of the system...}\\

However epistemic nature of quantum mechanics is not exactly same as the information-theoretic approach. The epistemic nature of the wavefunction requires the existence of ontic variables, however the information-theoretic approach relies on the fact that the act of measurement is about gaining the information about the system. As was clarified by Zurek \cite{4},\\
\\{\it
...Quantum measurements are usually analyzed in abstract terms of wavefunctions and Hamiltonians. Only very few discussions of the measurement
problem in quantum theory make an explicit effort to consider the crucial issue—the transfer of information. Yet obtaining knowledge is the very reason for making a measurement...}
\\

In 1999 \cite{2} Zeilinger formulated the {\it Foundational principle} for the information-theoretic approach to quantum foundations. The principles are given by\\
\\
$\bullet$ {\it An elementary system represents the truth value of one proposition.}\\
\\
$\bullet${\it An elementary system carries one bit of information.}\\
\\
Even though the principles formed a starting point for the rigorous development of the subject, it mostly remained as philosophical construct. These ideas were further refined by Peres \cite{9}, Fuchs and Peres \cite{11}, Fuchs \cite{10}, Mermin \cite{12}, and Wheeler \cite{13}. 

Entropy quantifies how much information we gain about a system on an average or to quantify the resources needed to store information \cite{5},\cite{6}. For a rigorous information-theoretic approach it is important to find relations among the quantum mechanical observables and entropy. One conceptual difficulty for this approach is that observables evolve unitarily. So all the observable values change with time. However entropy is invariant under a unitary transformation. Therefore no equality relations should be expected, but some inequalities might be useful to relate the concepts of quantum information and quantum mechanics.

In this paper we will find the upper and lower bounds for the expectation value of an observable and suprisingly those bounds depend on the entropy. We also find the upper bound for the uncertainty of an observable, which also depends on the entropy. Entropy is also defined as the uncertainty in the state of the physical system \cite{5}. We find a uncertainty like relation between the uncertainty of an observable and the entropy and show that the limit depends on the dimensions of the system. Also we find a reverse uncertainty relation which gives an upper bound for the product and sum of the variances of n quantum mechanical observables.
\section{The bounds for the expectation value of an observable A {\bf ($\langle A \rangle_\rho$)}}
The entropy is mathematically given by
\begin{equation}
S=-Tr(\rho log_2\rho)
\end{equation}
It is trivial to show that [\ref{appendix}]
\begin{equation}\label{eq1}
S \geq Tr(\rho-\rho^2) 
\end{equation}
{\bf Theorem 1.} {\it The expectation value of an hermitian operator $\hat{A}$, is bounded by the sum of the entropy and the purity of the state.}
\\
\\
{\it Proof.} Using the inequality [\ref{eq1}] 
\begin{equation}
Tr(|\hat{A}|)S \geq Tr(|\hat{A}|)Tr(\rho-\rho^2)
\end{equation}
\\
Since $\hat{A}$ is hermitian opeartor, it can be diagonalised and has real eigenvalues. So we take the diagonalised $\hat{A}$. (From here onwards we will always consider that $\hat{A}$ is diagonalised.)

We know that
\begin{equation}\label{eq4}
|Tr(DA)| \leq Tr(|D|)Tr(|A|)
\end{equation}
where D is diagonal matrix and A is any matrix. The $|A|$ represents the matrix with the absolute values of the elements of the matrix $A$.

Since the diagonal elements of $\rho$ are all greater than or equal to 0. The density matrix $\rho$ is given in the basis which diagonalises $\hat{A}$. Therefore from equ.[\ref{eq4}]
\\
\begin{equation}
\begin{split}
Tr(|\hat{A}|)(S+Tr(\rho^2))\geq Tr(|\hat{A}|)Tr(\rho) \geq  |Tr(\hat{A}\rho)|\\
\implies Tr(|\hat{A}|)(S+Tr(\rho^2)) \geq  |\langle A \rangle_\rho|
\end{split}
\end{equation}
Therfore
\begin{equation}
Tr(|\hat{A}|)(S+Tr(\rho^2)) \geq  \langle A \rangle_\rho \geq -Tr(|\hat{A}|)(S+Tr(\rho^2))
\end{equation}
This completes the proof.

The factor $Tr(|\hat{A}|)$ can be defined as a normalization factor as this factor is a constant for a particular operator. As we can see the expectation value is bounded from above and below by the sum of entropy and the purity of the state. 

\section{The bounds for the uncertainty in the value of an observable A  ($\sigma_{A}$)}

{\bf Theorem 2.} {\it The variance of an hermitian operator $\hat{A}$ whose all eigenvalues are non-negative, is bounded by the dimensions of the operator.}
\\
\\
{\it Proof.} From quantum theory \cite{7} we know that
\begin{equation}\label{eq7}
\sigma_{A}^{2} = Tr(\hat{A}^2\rho)-(Tr(\hat{A}\rho))^2
\end{equation}
From the inequality [\ref{eq4}]
\begin{equation}
Tr(\hat{A}^2\rho) \leq Tr(\hat{A})Tr(\hat{A}\rho)
\end{equation}
Therefore
\begin{equation}
Tr(\hat{A}^2\rho)-(Tr(\hat{A}\rho))^2\leq Tr(\hat{A})Tr(\hat{A}\rho)-(Tr(\hat{A}\rho))^2
\end{equation}
Now from equ. [\ref{eq7}]
\begin{equation}\begin{split}
\sigma_{A}^{2} \leq Tr(\hat{A}\rho)(Tr(\hat{A})-(Tr(\hat{A}\rho)))
\\
\implies \sigma_{A}^{2} \leq Tr(\hat{A}\rho)(Tr(\hat{A}(I-\rho)))
\end{split}
\end{equation}
Again using eq. [\ref{eq4}]
\begin{equation}
Tr(\hat{A}(I-\rho)) \leq  Tr(\hat{A})Tr(I-\rho) = Tr(\hat{A})(d-1)
\end{equation}
where d is the dimension of the operator $\hat{A}$. From equ. [\ref{eq7}], we have
\begin{equation}
\sigma_{A}^{2} \leq Tr(\hat{A}\rho)(Tr(\hat{A}(I-\rho)))\leq Tr(\hat{A}\rho) Tr(\hat{A})(d-1)
\end{equation}
Therefore
\begin{equation}
\sigma_{A}^{2} \leq \langle A \rangle_\rho Tr(\hat{A})(d-1)
\end{equation}
This completes the proof.

As the above result shows that the variance of any quantum mechanical hermitian opertors whose eigenvalues are positive is bounded above by the product of the dimensions of the system and the the expectation value of the observable with a normalization factor given by $Tr(\hat{A})$. 
\\
\\
{\bf Theorem 3.} {\it The variance of any general hermitian operator $\hat{A}$, is bounded by the sum of the entropy and the purity of the state.}
\\
\\
{\it Proof.} Since the diagonal elemnents of $\hat{A}^2$ are greater than or equal to 0.
\\
Therefore
\begin{equation}
Tr(\hat{A}^2)S \geq Tr(\hat{A}^2)Tr(\rho-\rho^2)
\end{equation}
Using eq. [\ref{eq4}]
\begin{equation}
Tr(\hat{A}^2)(S+Tr(\rho^2)) \geq Tr(\hat{A}^2)Tr(\rho) \geq Tr(\hat{A}^2\rho)
\end{equation}
From eq. [\ref{eq7}]
\begin{equation}
Tr(\hat{A}^2)(S+Tr(\rho^2))\geq \sigma_{A}^{2} + (Tr(\hat{A}\rho))^2
\end{equation}
Therfore
\begin{equation}
Tr(\hat{A}^2)(S+Tr(\rho^2))\geq \sigma_{A}^{2} + \langle A \rangle_\rho^2
\end{equation}
Since $\langle A \rangle_\rho^2$ is always greater than or equal to 0. Therefore
\begin{equation}\label{eq18}
Tr(\hat{A}^2)(S+Tr(\rho^2))\geq \sigma_{A}^{2}
\end{equation}
This completes the proof.

As shown above that the variance of any general quantum mechanical hermitian operator is bounded above by the sum of the entropy and the purity with a normalization factor given by $Tr(\hat{A}^2)$.\\
\\
{\bf Corollary 1.} {\it A weaker limit for the product of entropy $S$ and the uncertainty of an observable $\sigma_{A}$ which resembles Heisenberg's uncertainty relation, is given by
\begin{equation}
\Delta A\ S\leq f(d)\sqrt{Tr(\hat{A}^2)}
\end{equation} 
where $f(d)$ is a function dependent on the dimension $d$ given by $f(d)=log_2d\ \sqrt{log_2 2d}$.}\\
\\
{\it Proof.} From eq. [\ref{eq18}]
\begin{equation}
\sigma_{A}\leq\sqrt{Tr(\hat{A}^2)(S+Tr(\rho^2))} 
\end{equation}
Since $S\geq 0$,
\begin{equation}
\sigma_{A}S\leq S\sqrt{Tr(\hat{A}^2)(S+Tr(\rho^2))} 
\end{equation}
Since $S\leq log_2 d$ and $Tr(\rho^2)\leq 1$,
\begin{equation}
\sigma_{A}S\leq log_2 d\ \sqrt{log_2 d+1}\  \sqrt{Tr(\hat{A}^2)}
\end{equation}
This gives us the desired result
\begin{equation}
\Delta A\ S\leq f(d)\sqrt{Tr(\hat{A}^2)}
\end{equation}
where $f(d)$ is given by $f(d)=log_2d\ \sqrt{log_2 2d}$.\\
\\
This completes the proof.

The above result shows that the product of the standard deviation and the {\it von Neumann entropy} is bounded above by a function dependent only on the dimensions of the system with a normalization factor given by $\sqrt{Tr(\hat{A}^2)}$.

A recent result by A.K. Pati et. al \cite{14} showed the existence of reverse uncertainty relations. The reverese uncertainty results means that the product or the sum of the variances of the observables are also bounded from above. A similar reverse uncertainty relation can be obtained where the bound is dependent on the sum of the entropy and the purity of the state.\\
\\
{\bf Corollary 2.} {\it An upper bound for the product and sum of the variances of two observables which resembles the reverse Heisenberg's uncertainty relation, is given by}
\begin{equation}
\sigma_{A}^{2}\sigma_{B}^{2}\leq Tr(\hat{A}^2)Tr(\hat{B}^2)(S+Tr(\rho^2))^2
\end{equation} 
and
\begin{equation}
\sigma_{A}^{2}+\sigma_{B}^{2}\leq Tr(\hat{A}^2+\hat{B}^2)(S+Tr(\rho^2))
\end{equation} 
\\
{\it Proof.} From eq. [\ref{eq18}], we have
\begin{equation}\label{eq26}
\sigma_{A}^{2}\leq Tr(\hat{A}^2)(S+Tr(\rho^2))
\end{equation} 
Also 
\begin{equation}\label{eq27}
\sigma_{B}^{2}\leq Tr(\hat{B}^2)(S+Tr(\rho^2))^2
\end{equation}
Since variances are always positive, therefore multiplying the eq. [\ref{eq26}] and eq. [\ref{eq27}]
\begin{equation}
\sigma_{A}^{2}\sigma_{B}^{2}\leq Tr(\hat{A}^2)Tr(\hat{B}^2)(S+Tr(\rho^2))^2
\end{equation}
Similarly adding eq. [\ref{eq26}] and eq. [\ref{eq27}], we get
\begin{equation}
\sigma_{A}^{2}+\sigma_{B}^{2}\leq Tr(\hat{A}^2+\hat{B}^2)(S+Tr(\rho^2))
\end{equation}
The above result could be trivially extended to n quantum mechanical observables.
\begin{equation}
\sigma_{A_1}^{2}\sigma_{A_2}^{2}...\sigma_{A_n}^{2}\leq Tr(\hat{A_1}^2)Tr(\hat{A_2}^2)...Tr(\hat{A_n}^2)(S+Tr(\rho^2))^n
\end{equation}
and
\begin{equation}
\sigma_{A_1}^{2}+\sigma_{A_2}^{2}+...\sigma_{A_n}^{2}\leq Tr(\hat{A}^2+\hat{A_2}^2+...\hat{A_n}^2)(S+Tr(\rho^2))
\end{equation}
 
\section{Conclusions}
As we know that in an experiment on any quantum system, the expectation values of different observables and the uncertainty of those variables define the system. As shown above these quantities are bounded by the sum of the entropy and the purity of the state. The results point out that quantum information is intrinsically connected to quantum mechanics. Though {\it von Neumann entropy} is only dependent on the quantum state, any observable values which depends on the measurements carried out by the observer is bounded by the entropy. Therefore quantum information or {\it von Neumann entropy} can also be defined as the maximum value of the expectation value of any general quantum mechanical observable or the maximum uncertainty of any general quantum mechanical observable (keeping in mind that the purity has to be subtracted from these values and also the presence of the normalization constant as defined above). \\
\\
{\bf Acknowledgements}\\
\\
I would like to thank Chandan Datta, IOP Bhubaneswar for valuable suggestions and discussions towards this work.

\section{Appendix}
\label{appendix}
As 
\begin{equation}
S=-Tr(\rho log\rho)=-\langle log \rho \rangle
\end{equation}
Using the Mercartor series for logarithm, we have
\begin{equation}
-\langle log\rho \rangle=\langle 1-\rho \rangle+\langle (1-\rho)^2 \rangle/2+...
\end{equation}
Therefore
\begin{equation}
S=S_L + some \  positive\  terms
\implies
S\geq S_L
\end{equation}
\end{document}